\documentclass[conference]{IEEEtran}
\IEEEoverridecommandlockouts

\usepackage{cite}
\usepackage{amsmath,amssymb,amsfonts}
\usepackage{algorithmic}
\usepackage{graphicx}
\usepackage{textcomp}
\usepackage{xcolor}
\usepackage{booktabs}
\def\BibTeX{{\rm B\kern-.05em{\sc i\kern-.025em b}\kern-.08em
    T\kern-.1667em\lower.7ex\hbox{E}\kern-.125emX}}
\begin{document}

\title{No Free Lunch from Audio Pretraining in Bioacoustics: \\A Benchmark Study of Embeddings}
\author{\IEEEauthorblockN{Chenggang Chen}
\IEEEauthorblockA{\textit{Department of Biomedical Engineering} \\
\textit{Johns Hopkins University}\\
Baltimore, USA \\
ccgjhu1988@gmail.com}
\and
\IEEEauthorblockN{Zhiyu Yang}
\IEEEauthorblockA{\textit{Department of Biomedical Engineering} \\
\textit{Johns Hopkins University}\\
Baltimore, USA \\
yangzhiyuzyyy@gmail.com}}

\maketitle

\begin{abstract}
Bioacoustics, the study of animal sounds, offers a non-invasive method to monitor ecosystems. Extracting embeddings from audio-pretrained deep learning (DL) models without fine-tuning has become popular for obtaining bioacoustic features for tasks. However, a recent benchmark study reveals that while fine-tuned audio-pretrained VGG and transformer models achieve state-of-the-art performance in some tasks, they fail in others. This study benchmarks 11 DL models on the same tasks by reducing their learned embeddings' dimensionality and evaluating them through clustering. We found that audio-pretrained DL models 1) without fine-tuning even underperform fine-tuned AlexNet, 2) both with and without fine-tuning fail to separate the background from labeled sounds, but ResNet does, and 3) outperform other models when fewer background sounds are included during fine-tuning. This study underscores the necessity of fine-tuning audio-pretrained models and checking the embeddings after fine-tuning. Our codes are available: https://github.com/NeuroscienceAI/Audio\_Embeddings
\end{abstract}


\section{Introduction}
Computational bioacoustics with machine learning has been widely used for classifying and detecting animal sounds \cite{stowell2022computational}. The major challenge is to extract embeddings, the high-dimensional representation of audio features. There are at least three methods to obtain the embeddings: 1) Predefined acoustic features like mel frequency cepstrum coefficient (MFCC) \cite{xie2016acoustic, elie2016vocal, barker2021cultural, phillips2018revealing, clink2021unsupervised, hilasaca2021visualization}, 2) Spectrograms of single vocalizations or syllables \cite{sainburg2019parallels, sainburg2020finding, sainburg2021toward, robotka2023sparse}, and 3) Audio-pretrained models. Using audio-pretrained models without fine-tuning is popular \cite{coban2020transfer, sethi2022soundscapes, nolasco2022rank, mcginn2023feature, ghani2023global}. For example, \cite{sethi2020characterizing} uses YouTube audio-pretrained VGGish \cite{hershey2017cnn} to calculate fingerprints of soundscapes from a variety of ecosystems. \cite{sarkar23_interspeech} shows that eleven human speech-pretrained models could distinguish the identities of ten animals. Do we still need fine-tuning? \cite{lakdari2024mel} recently found that under noisy conditions, MFCC outperforms all three audio-pretrained models without fine-tuning. To the best of our knowledge, no studies compare the embeddings of animal sounds before and after fine-tuning the audio-pretrained models.

Fine-tuned audio-pretrained models are not considered state-of-the-art either. Two recent studies \cite{hagiwara2023beans, hagiwara2023aves} fine-tuned audio-pretrained models: one is VGGish, and the other is bioacoustics-pretrained transformer models (AVES-bio), and benchmarked them over ten datasets. In three classification datasets, both the audio- and bioacoustics-pretrained models outperform all other models, achieving more than 0.8 accuracy. In contrast, in three detection tasks, they exhibit less than 0.4 precision and even underperform compared to other DL models. The reason is unclear, partly attributed to the absence of benchmarked embeddings from audio-pretrained and other deep learning models.

Our contributions in bioacoustics are as follows:
\begin{enumerate}
\item First study benchmarks DL models with embeddings. 
\item First study evaluates 11 DL models on same datasets. 
\item Finds fine-tuning is necessary for pretrained models.
\item Explains why pretrained models fail on some datasets. 
\item Improves task performance of the pretrained models.
\end{enumerate}

\section{Methods}
\begin{figure}[htbp]
  \centering
  \includegraphics[width=\linewidth]{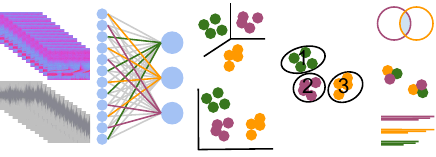}
  \caption{Spectrograms or waveforms from ten datasets are fed into eleven deep neural networks to learn the audio embeddings. Embeddings (high-dimensional and latent) are projected (t-SNE and UMAP) to two dimensions for visualization and clustering (KMeans). Clustering performance is evaluated with three metrics (NMI, ARI, and Silhouette coefficient).}
  \label{fig:Fig1_diagram}
\end{figure}

Figure~\ref{fig:Fig1_diagram} shows the four major steps of this study: learning embeddings, dimensionality reduction, clustering, and evaluation.
\subsection{Eleven deep learning models}
\begin{figure*}[ht]
  \centering
  \includegraphics[width=\textwidth]{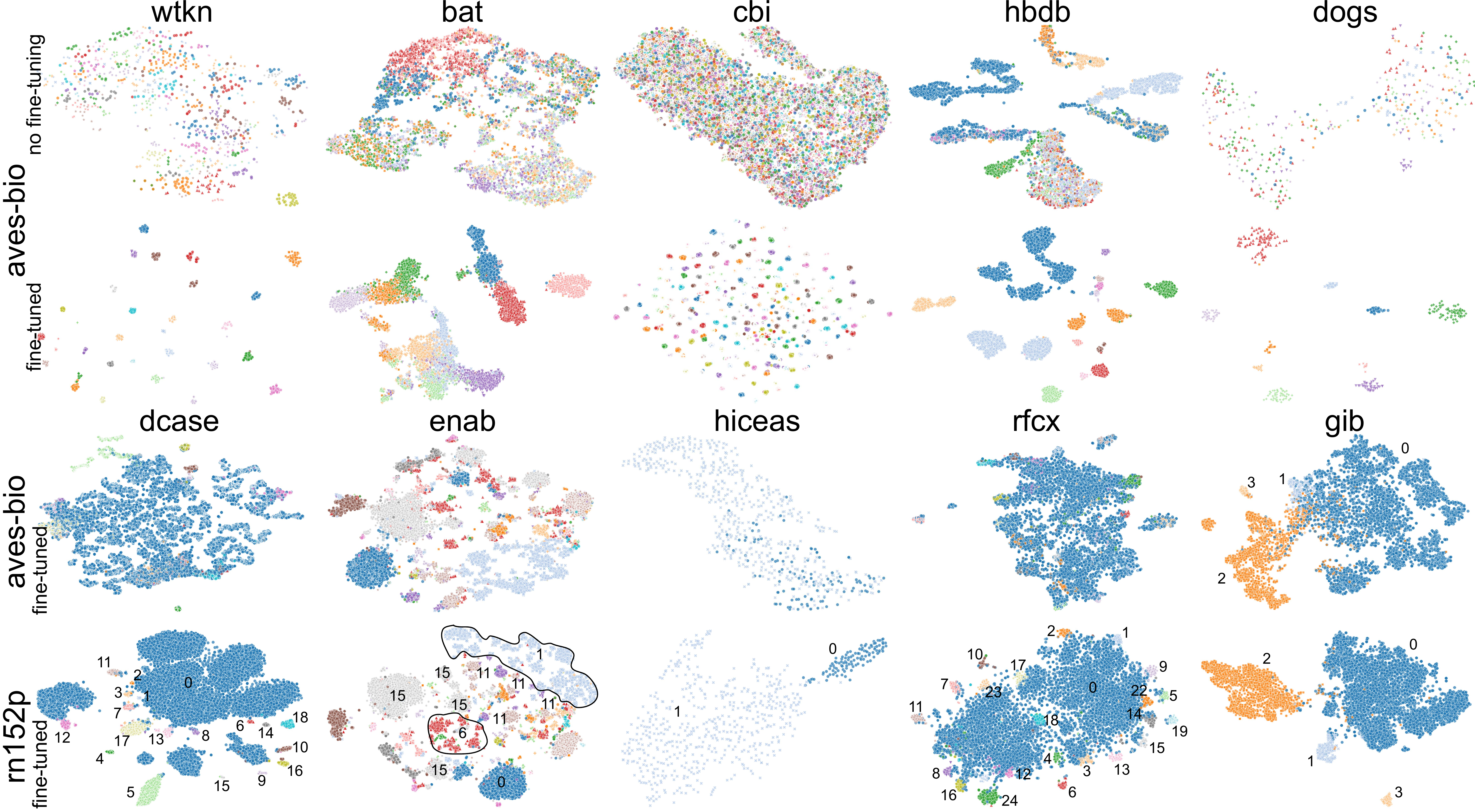}
  \caption{High dimensional embeddings of five classification (row 1st, 2nd) and five detection (row 3rd, 4th) datasets are projected to two dimensions using t-SNE (perplexity=50, iteration=5000). First row shows the embeddings extracted from bioacoustics-pretrained model without fine-tuning. Deep blue dots (class 0, same as Figure 3) in the 3rd and 4th rows are the unlabeled background sounds.}
  \label{fig:Fig2_tSNE_whole}
\end{figure*}

We trained deep learning models using PyTorch on a single GPU (RTX A5000, NVIDIA). We trained three ResNet (rn18, rn50, rn152) models from scratch and fine-tuned five ImageNet-pretrained AlexNet (alexp), ResNet (rn18p, rn50p, rn152p), and Swin Transformer (swinTp) models, one YouTube audio-pretrained VGG (vggish\cite{hershey2017cnn}) model, and two audio (all sounds and animal sounds) pretrained HuBERT model (aves\cite{hagiwara2023aves}). We loaded the weights of alexnet, resnet and swinTp from Torchvision, vggish from PyTorch Hub ('harritaylor/torchvggish'), and aves that ported to Torchaudio from Github (earthspecies/aves). AVES models are fine-tuned with raw waveform and other models are trained or fine-tuned with spectrogram. Notice that the original BEANS only has resnet and vggish, and we integrated the alexp, swinTp, and two aves models into the BEANS.

We used the same optimizer setting (Adam), batch size (32), and learning rates (0.00001, 0.00005, 0.0001) as BEANS. We picked the trained model that has the best performance on the validation dataset. The hyperparameters search (i.e., three learning rates), training, and validation have never been used on the test datasets.

\subsection{Embeddings dimensionality reduction}
We used t-SNE \cite{maaten2008visualizing} (Figure~\ref{fig:Fig2_tSNE_whole}) and UMAP \cite{mcinnes2018umap} to reduce the high dimensional embeddings. In t-SNE, the primary parameter is perplexity, and we tested three values: 25, 50, and 100. Since higher iterations number result in better performance, we used 5000 (default is 1000). In UMAP, the primary parameters are number of neighbor and minimal distance, and we tested two values for each of them (50, 100; 0.1, 0.5). 
\begin{table*}[ht]
  \caption{Clustering (KMeans, k=class number) of two dimensional learnt embeddings (similar to Figure 2; t-SNE, perplexity=25, iteration=5000) is evaluated with normalized mutual information (NMI) on 11 models and 10 datasets. Larger values mean better clustering. The best and second best data are highlighted and underlined, respectively. Last six lines show data from untrained (1), pretrained without fine-tuning (2-4), and incomplete fine-tuned (5-6) models .}
  \label{tab:tSNE whole datasets}
  \centering
  \begin{tabular}{@{}l l l l l l l l l l l@{}}
    \toprule
    \textbf{Model} & 
    \textbf{wtkn} & \textbf{bat} & \textbf{cbi} & \textbf{hbdb} & \textbf{dogs} 
    & \textbf{dcase}& \textbf{enab}& \textbf{hiceas} & \textbf{rfcx} & \textbf{gib}\\
    \midrule
    AlexNet image-pretrained	&0.551	&0.298	&0.517	&0.314	&0.367	&0.150	&0.346	&0.003	&0.081	&0.114\\
    ResNet18	&0.921	&0.345	&0.722	&0.544	&0.937	&0.216	&0.584	&0.219	&0.179	&0.414\\
    ResNet50	&0.836	&0.350	&0.847	&0.548	&0.570	&0.223	&0.563	&\underline{0.271}	&0.203	&0.422\\
    ResNet152	&0.817	&0.566	&0.778	&0.477	&0.482	&0.227	&0.577	&0.021	&0.207	&0.397\\
    ResNet18 image-pretrained	&0.909	&0.331	&0.901	&0.453	&0.886	&0.221	&0.590	&0.158	&0.195	&\underline{0.513}\\
    ResNet50 image-pretrained	&0.895	&0.446	&0.839	&0.490	&0.814	&\underline{0.303}	&\underline{0.596}	&0.257	&\underline{0.275}	&\textbf{0.519}\\
    ResNet152 image-pretrained	&0.919	&0.520	&0.889	&0.612	&0.760	&\textbf{0.324}	&0.579	&\textbf{0.405}	&\textbf{0.295}	&0.347\\
    Swin Transformer image-pretrained	&0.825	&0.516	&0.818	&0.471	&0.714	&0.212	&0.571	&0.011	&0.114	&0.179\\
    VGGish audio-pretrained	&0.919	&\textbf{0.830}	&0.828	&\textbf{0.739}	&0.845	&0.293	&\textbf{0.599}	&0.104	&0.204	&0.314\\
    AVES audio-pretrained	&\underline{0.963}	&\underline{0.775}	&\underline{0.950}	&0.733	&\textbf{1.000}	&0.223	&0.586	&0.145	&0.252	&0.370\\
    AVES bioacoustics-pretrained	&\textbf{0.971}	&0.653	&\textbf{0.981}	&\textbf{0.776}	&\textbf{1.000}	&0.223	&0.568	&0.146	&0.253	&0.410\\
    \midrule
    ResNet152 no pre-training no fine-tuning &0.206&0.049&0.304&0.088&0.109&0.038&0.029&0.006&0.047&0.001\\
    ResNet152 image-pretrained no fine-tuning &0.276&0.092&0.340&0.159&0.150&0.092&0.059&0.010&0.059&0.002\\
    VGGish audio-pretrained no fine-tuning    &0.516&0.142&0.339&0.442&0.163&0.145&0.164&0.011&0.095&0.015\\
    AVES bioacous.-pretrained no fine-tuning	&0.581&0.215&0.373&0.487&0.195&0.147&0.287&0.000&0.121&0.013\\
    AVES bioacous.-pretrained 1 epoch	 &0.530&0.249&0.354&0.431&0.167&0.108&0.173&0.008&0.024&0.063\\
    AVES bioacous.-pretrained 3 epochs	&0.629&0.467&0.583&0.553&0.287&0.227&0.382&0.170&0.054&0.286\\
    \bottomrule
  \end{tabular}
\end{table*}
\subsection{Ten animal sounds datasets}
\begin{figure}[ht]
  \centering
  \includegraphics[width=\linewidth]{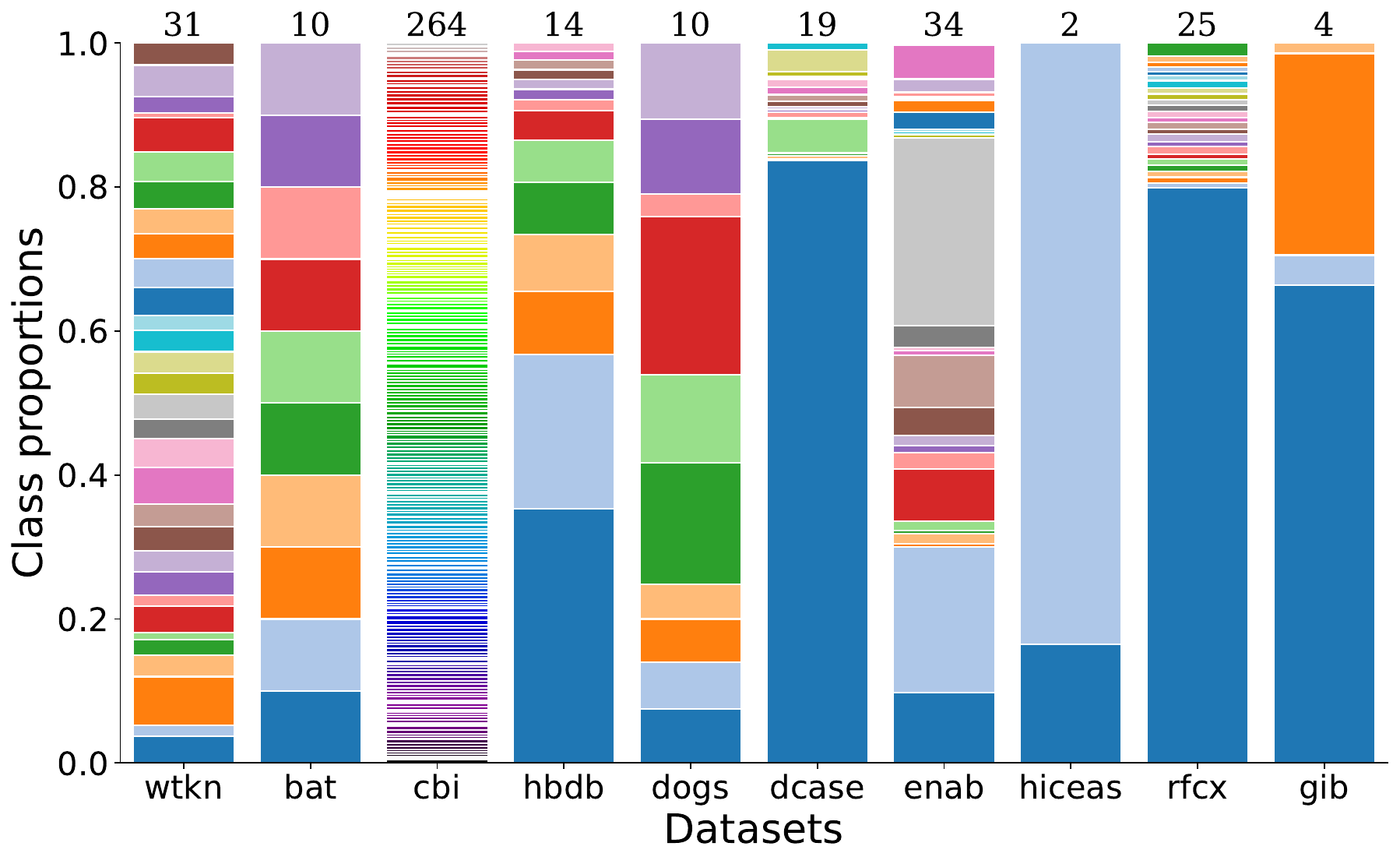}
  \caption{Unbalanced number of samples in each class. The last five datasets (dcase to gib) are used for recognition task where the first class (deep blue bars at bottom, same as Figure 2) is the background sounds without human labeling or annotation.}
  \label{fig:Fig3_class_num}
\end{figure}

\begin{figure}[ht]
  \centering
  \includegraphics[width=\linewidth]{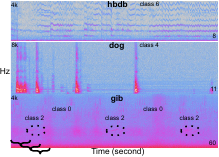}
  \caption{Spectrogram of three example audios. Each audio of hbdb and dog dataset (classification task) only contains one class. Each audio of gib dataset (detection task) contains multiple chunks belonging to different classes (three class 2 and others are class 0). Braces: 4 seconds of sliding window.}
  \label{fig:Fig4_audio_example}
\end{figure}

We train or fine-tune models on datasets used by BEANS \cite{hagiwara2023beans}(Figure~\ref{fig:Fig3_class_num}). We used the same codes for pre-processing the datasets. 
\begin{itemize}
\item wtkn: Watkins marine mammal sound database. 
\item bats: Egyptian fruit bat calls and class is caller identity. 
\item cbi: Cornell bird identification competition data. 
\item dogs: Domestic dogs' barks in different situations. 
\item hbdb: Wild and cultured mosquito wing beat sounds. 
\item dcase: Mammal and bird calls with annotations. 
\item enab: Eastern north American birds dawn chorus. 
\item hiceas: Hawaiian islands cetacean and ecosystem assessment survey of minke whale “boing” call. 
\item rfcx: Rain forest connection includes frog and bird. 
\item gib: Hainan gibbon calls annotated with times and types. 
\end{itemize}
The first five datasets are used for classification, and the last five for detection. For classification (top and middle, Figure~\ref{fig:Fig4_audio_example}), the duration of each audio file is adjusted to meet the minimum duration requirement of each dataset (3, 5, 10, 3, and 10 seconds) through either padding with silence at the end or truncation. For detection (bottom, Figure~\ref{fig:Fig4_audio_example}), each raw audio file is chunked into specific durations (2, 2, 10, 10, and 4 seconds) using a sliding window technique, with each window's overlap set at half of its duration (braces, Figure~\ref{fig:Fig4_audio_example}). A non-zero class is detected if the overlap between the sliding window and any annotation exceeds 20 percent; otherwise, this window is labeled as class 0.

\section{Experiments}
\subsection{Audio-pretrained models need fine-tuning}
Figure~\ref{fig:Fig2_tSNE_whole} displays the embeddings extracted from bioacoustics-pretrained AVES and image-pretrained ResNet152 after t-SNE dimensionality reduction. In the five classification datasets (1st and 2nd row), the bioacoustics-pretrained model struggles to group animal sounds from the same class together without fine-tuning, with the exception of the hbdb dataset, which shows clustered samples belonging to the same class. Figure~\ref{fig:Fig4_audio_example} explains this phenomenon. It is attributed to the fact that hbdb audio is filled with animal sounds, whereas dog audio contains less than 30 percent. This observation aligns with studies that extract embeddings without fine-tuning using precisely trimmed audios \cite{thomas2022practical}.

Next, we quantitatively evaluate the clustering performance using normalized mutual information (NMI), where 1 indicates that every sample from its own class is correctly clustered without any error. ResNet152, without pre-training and fine-tuning, works similarly to previous studies that use spectrograms as embeddings \cite{sainburg2020finding}, and its NMI is lower than that of pretrained models (Table~\ref{tab:tSNE whole datasets}). Without fine-tuning, the bioacoustics-pretrained model performs better than the audio-pretrained model, and both significantly outperform image-pretrained models. However, both bioacoustics- and audio-pretrained models without fine-tuning underperform fine-tuned image-pretrained AlexNet in seven datasets. They outperform AlexNet by a large margin only for the hbdb dataset, consistent with Figure~\ref{fig:Fig2_tSNE_whole} and ~\ref{fig:Fig4_audio_example}. Notably, the NMIs of both models are lower than those of all other DL models across all datasets. After one epoch of fine-tuning, only three datasets show a larger NMI than the non-tuned baseline, likely due to a mismatch between the pretraining and fine-tuning tasks. After three epochs, nine datasets exhibit a larger NMI than the baseline. In fully tuned models, the NMI increases by at least 1.5 times.

\subsection{Audio-pretrained models are not the best}
In the five classification datasets, both the best and second-best results are achieved using audio-pretrained models (Table~\ref{tab:tSNE whole datasets}). The clustering performance measured here aligns with accuracy \cite{hagiwara2023aves}: the dog dataset, which has the highest accuracy, also exhibits the largest NMI. Conversely, in the four detection datasets (dcase, hiceas, rfcx, and gib, Table~\ref{tab:tSNE whole datasets}), neither the best nor the second-best results are achieved with audio-pretrained models. Instead, the image-pretrained rn152p model secures the best results in three datasets.

Fine-tuned bioacoustics-pretrained model (aves-bio) intermingles labeled sounds with backgrounds (3rd row, Figure~\ref{fig:Fig2_tSNE_whole}). A significant challenge is the vast number of background sounds (deep blue dots, Figure~\ref{fig:Fig2_tSNE_whole}), which constitute uniform noise lacking distinct acoustic features, as well as animal sounds from other species (gib, Figure~\ref{fig:Fig4_audio_example}). Consequently, background sounds in the latent embedding space are either widely distributed (representing flat noise) or form distinct clusters (representing sounds from other species). By contrast, the image-pretrained model (rn152p) successfully tackles this challenge by constraining the embeddings of background sounds while distinctly separating the embeddings of labeled sounds (4th row, Figure~\ref{fig:Fig2_tSNE_whole}). However, its performance is not uniformly excellent across all datasets. For instance, in the enab dataset, samples belonging to classes 1 and 6 form loose clusters, while classes 11 and 15 are represented by four separate clusters.
\subsection{Control experiments}
Testing t-SNE with a perplexity of 100 (versus 25 in Table~\ref{tab:tSNE whole datasets}) and UMAP showed that rn152p continues to outperform audio-pretrained models (Table~\ref{tab:p100&UMAP}). Further experiment with two key UMAP parameters yielded similar outcomes (Supplementary material). Beyond NMI, we evaluated two other popular clustering metrics, obtaining consistent results (Table~\ref{tab:ARI&Silho}).

\begin{table}[ht]
  \caption{Same clustering evaluation metric (NMI) as Table 1 but different dimensionality reduction parameter (t-SNE, perplexity=100) and method (UMAP, neighbors=50, min dist=0.5).}
  \label{tab:p100&UMAP}
  \centering
  \begin{tabular}{@{}lllllll@{}}
    \toprule
    \multicolumn{1}{c}{} & \multicolumn{2}{c}{\textbf{dcase}}& \multicolumn{2}{c}{\textbf{rfcx}} &\multicolumn{2}{c}{\textbf{gib}}\\ \hline
    \textbf{Model} & 
    p100 & umap & p100 & umap & p100 & umap\\
    \midrule
    alexp &0.149&	0.153&	0.081	&0.077	&0.136	&0.137\\
    rn18&0.215	&0.217	&0.184	&0.176	&0.436	&0.417\\
    rn50&0.225	&0.227	&0.203	&0.196	&0.412	&0.406\\
    rn152&0.228	&0.229	&0.211	&0.200	&0.398	&0.413\\
    rn18p&0.221	&0.225	&0.198	&0.179	&\underline{0.496}	&\underline{0.488}\\
    rn50p & 0.297	& \underline{0.300} & \underline{0.272} &	0.258	&\textbf{0.543}	& \textbf{0.510}\\
    rn152p &\underline{0.305}	&\textbf{0.329}	&\textbf{0.292}	&\textbf{0.281}	&0.360	&0.355\\
    swinTp &0.212	&0.289	&0.118	&0.196	&0.217	&0.349\\
    vggish & \textbf{0.311}	&0.204	&0.203	&0.109	&0.325	&0.304\\
    aves-all&0.249	&0.251	&0.248	&0.254	&0.437	&0.447\\
    aves-bio&0.249	&0.262	&0.258	&\underline{0.261}	&0.416	&0.419\\
    \bottomrule
  \end{tabular}
\end{table}
\begin{table}[ht]
  \caption{Other metrics. Larger values indicate better clustering. ARI: adjusted random index, Silho: Silhouette Coefficient.}
  \label{tab:ARI&Silho}
  \centering
  \begin{tabular}{@{}lllllll@{}}
    \toprule
    \multicolumn{1}{c}{} & \multicolumn{2}{c}{\textbf{dcase}}& \multicolumn{2}{c}{\textbf{rfcx}} &\multicolumn{2}{c}{\textbf{gib}}\\ \hline
    \textbf{Model} & 
    ARI & Silho & ARI & Silho & ARI & Silho\\
    \midrule
    alexp	&0.012	&0.090&	0.001&	0.033&	0.107	&\textbf{0.217}\\
    rn18	&0.019	&\underline{0.141}	&0.002	&0.063	&0.298	&0.156\\
    rn50	&0.019	&\textbf{0.172}	&0.004	&0.061	&0.323	&\underline{0.211}\\
    rn152	&0.020	&\textbf{0.172}	&0.005	&0.068	&0.279	&0.174\\
    rn18p	&0.021	&0.086	&0.006	&0.007	&\underline{0.344}	&0.140\\
    rn50p	&\underline{0.033}	&0.083	&\underline{0.017}	&-0.00	&\textbf{0.353}	&0.144\\
    rn152p	&\textbf{0.035}	&0.078	&\textbf{0.020}	&-0.00	&0.271	&0.096\\
    swinTp	&0.020	&0.111	&0.001	&-0.03	&0.143	&0.016\\
    vggish	&0.029	&0.045	&0.005	&0.074	&0.249	&0.181\\ 
    aves-all	&0.018	&0.018	&0.009	&\textbf{0.170}	&0.286	&0.063\\
    aves-bio	&0.016	&0.029	&0.010	&\underline{0.118}	&0.301	&0.106\\
    \bottomrule
  \end{tabular}
\end{table}

\subsection{Data cleaning boost model performance}
Audio-pretrained models struggled with three detection tasks, failing to distinguish between background and labeled sounds (Figure~\ref{fig:Fig2_tSNE_whole}). We hypothesized that removing audio files without labels (from training, validation, and testing sets) would reduce sample imbalance and condense the background's latent space. As anticipated, data cleaning enhanced performance (Table~\ref{tab:datacleaning}).

\begin{table}[ht]
  \caption{Data cleaning guided by embeddings substantially and consistently improves average precision for detection tasks.} 
  \label{tab:datacleaning}
  \centering
  \begin{tabular}{@{}lllllll@{}}
    \toprule
    \multicolumn{1}{c}{} & \multicolumn{2}{c}{\textbf{dcase}}& \multicolumn{2}{c}{\textbf{rfcx}} &\multicolumn{2}{c}{\textbf{gib}}\\ \hline
    \textbf{Model} & 
    beans & ours & beans & ours & beans & ours\\
    \midrule
    rn18 &      $0.161$&$0.179$      & $0.064$&$0.106$       & $0.164$&$0.326$\\
    rn50 &      $0.183$&$0.169$      & $0.055$&$0.116$       & $0.215$&$0.297$\\
    rn152 &     $0.154$&$0.165$      & $0.069$&$0.093$       & $0.248$&$0.297$\\
    rn18p &     $0.223$&$0.247$      & $0.079$&$0.122$       & \textbf{0.316} & \underline{0.438}\\
    rn50p &     $0.178$&$0.197$      & $0.087$&$0.123$       & $0.155$&$0.340$\\
    rn152p &    $0.198$&$0.233$      & $0.085$&$0.111$       & $0.230$&$0.392$\\
    vggish & \underline{0.335} & \underline{0.424} & \textbf{0.140} & \textbf{0.218} & $0.150$&$0.318$\\
    aves-bio & \textbf{0.392} & \textbf{0.452}  & \underline{0.130} & \underline{0.171}       & \underline{0.284} &\textbf{0.457}\\
    \bottomrule
  \end{tabular}
\end{table}

\section{Discussion}


This paper sets out to achieve three objectives. First, it benchmarks embeddings to uncover the representational learning capabilities of DL models. While accuracy, defined as the ratio of correctly predicted samples to all samples, overlooks background sounds, latent embeddings reveal that the challenge in detection tasks lies in differentiating labeled sounds from background. Models or data preprocessing capable of doing this achieve higher accuracy. Beyond providing insights for model improvement, this analysis also offers ecological insights, allowing us to discern the proximity or distance between species in the embedding space and identify species that may encompass subspecies.

Second, it compares audio-pretrained models without fine-tuning against fine-tuned models across ten datasets. Our findings underscore the necessity of fine-tuning, even for models pretrained on animal sounds. The distinction between our study and previous ones, which extract embeddings without fine-tuning, is attributed to the datasets used. Studies foregoing fine-tuning typically rely on clearly segmented audio of high signal quality, some collected in anechoic chambers \cite{robotka2023sparse, sarkar23_interspeech, zhang2018automatic, sainburg2020finding}. However, real-world scenarios often involve background noise, making fine-tuned pretrained models the superior choice for embedding extraction.

Third, this paper explores why audio-pretrained models fail on some datasets. Two main reasons emerge: one is the inability of audio-pretrained models to adapt to background sounds lacking distinct features, whereas image-pretrained models show greater suitability. Another reason is attributed to the network structure, with both image-pretrained AlexNet and Swin Transformer under performing compared to ResNet.

This study has its limitations. First, it removes audio files without any labeled sounds from detection datasets. While this approach is straightforward and easy to implement, it aids classification but still requires a human or an algorithm to determine the presence of target animal sound in the audio. Second, the datasets used in this study were collected in a variety of acoustic environments (ocean, rain forest, lab, home, etc.). We recommend that future research assess their models and methods (e.g., data augmentation \cite{padovese2021data, nanni2020data}) over a broader range of datasets, taking both accuracy and embeddings into consideration. Finally, this study solves the class imbalance issue by removing background samples. More advanced methods should be considered \cite{yang2021delving, chen2024neural}.

\bibliographystyle{IEEEtran}
\bibliography{FreeLunch}

\begin{thebibliography}{10}
\providecommand{\url}[1]{#1}
\csname url@samestyle\endcsname
\providecommand{\newblock}{\relax}
\providecommand{\bibinfo}[2]{#2}
\providecommand{\BIBentrySTDinterwordspacing}{\spaceskip=0pt\relax}
\providecommand{\BIBentryALTinterwordstretchfactor}{4}
\providecommand{\BIBentryALTinterwordspacing}{\spaceskip=\fontdimen2\font plus
\BIBentryALTinterwordstretchfactor\fontdimen3\font minus \fontdimen4\font\relax}
\providecommand{\BIBforeignlanguage}[2]{{%
\expandafter\ifx\csname l@#1\endcsname\relax
\typeout{** WARNING: IEEEtran.bst: No hyphenation pattern has been}%
\typeout{** loaded for the language `#1'. Using the pattern for}%
\typeout{** the default language instead.}%
\else
\language=\csname l@#1\endcsname
\fi
#2}}
\providecommand{\BIBdecl}{\relax}
\BIBdecl

\bibitem{stowell2022computational}
D.~Stowell, ``Computational bioacoustics with deep learning: a review and roadmap,'' \emph{PeerJ}, vol.~10, p. e13152, 2022.

\bibitem{xie2016acoustic}
J.~Xie, M.~Towsey, J.~Zhang, and P.~Roe, ``Acoustic classification of australian frogs based on enhanced features and machine learning algorithms,'' \emph{Applied Acoustics}, vol. 113, pp. 193--201, 2016.

\bibitem{elie2016vocal}
J.~E. Elie and F.~E. Theunissen, ``The vocal repertoire of the domesticated zebra finch: a data-driven approach to decipher the information-bearing acoustic features of communication signals,'' \emph{Animal cognition}, vol.~19, pp. 285--315, 2016.

\bibitem{barker2021cultural}
A.~J. Barker, G.~Veviurko, N.~C. Bennett, D.~W. Hart, L.~Mograby, and G.~R. Lewin, ``Cultural transmission of vocal dialect in the naked mole-rat,'' \emph{Science}, vol. 371, no. 6528, pp. 503--507, 2021.

\bibitem{phillips2018revealing}
Y.~F. Phillips, M.~Towsey, and P.~Roe, ``Revealing the ecological content of long-duration audio-recordings of the environment through clustering and visualisation,'' \emph{PloS one}, vol.~13, no.~3, p. e0193345, 2018.

\bibitem{clink2021unsupervised}
D.~J. Clink and H.~Klinck, ``Unsupervised acoustic classification of individual gibbon females and the implications for passive acoustic monitoring,'' \emph{Methods in Ecology and Evolution}, vol.~12, no.~2, pp. 328--341, 2021.

\bibitem{hilasaca2021visualization}
L.~M.~H. Hilasaca, L.~P. Gaspar, M.~C. Ribeiro, and R.~Minghim, ``Visualization and categorization of ecological acoustic events based on discriminant features,'' \emph{Ecological Indicators}, vol. 126, p. 107316, 2021.

\bibitem{sainburg2019parallels}
T.~Sainburg, B.~Theilman, M.~Thielk, and T.~Q. Gentner, ``Parallels in the sequential organization of birdsong and human speech,'' \emph{Nature communications}, vol.~10, no.~1, p. 3636, 2019.

\bibitem{sainburg2020finding}
T.~Sainburg, M.~Thielk, and T.~Q. Gentner, ``Finding, visualizing, and quantifying latent structure across diverse animal vocal repertoires,'' \emph{PLoS computational biology}, vol.~16, no.~10, p. e1008228, 2020.

\bibitem{sainburg2021toward}
T.~Sainburg and T.~Q. Gentner, ``Toward a computational neuroethology of vocal communication: from bioacoustics to neurophysiology, emerging tools and future directions,'' \emph{Frontiers in Behavioral Neuroscience}, vol.~15, p. 811737, 2021.

\bibitem{robotka2023sparse}
H.~Robotka, L.~Thomas, K.~Yu, W.~Wood, J.~Elie, M.~Gahr, and F.~E. Theunissen, ``Sparse ensemble neural code for a complete vocal repertoire,'' \emph{Cell Reports}, vol.~42, no.~2, 2023.

\bibitem{coban2020transfer}
E.~B. Coban, D.~Pir, R.~So, and M.~I. Mandel, ``Transfer learning from youtube soundtracks to tag arctic ecoacoustic recordings,'' in \emph{ICASSP 2020-2020 IEEE International Conference on Acoustics, Speech and Signal Processing (ICASSP)}.\hskip 1em plus 0.5em minus 0.4em\relax IEEE, 2020, pp. 726--730.

\bibitem{sethi2022soundscapes}
S.~S. Sethi, R.~M. Ewers, N.~S. Jones, J.~Sleutel, A.~Shabrani, N.~Zulkifli, and L.~Picinali, ``Soundscapes predict species occurrence in tropical forests,'' \emph{Oikos}, vol. 2022, no.~3, p. e08525, 2022.

\bibitem{nolasco2022rank}
I.~Nolasco and D.~Stowell, ``Rank-based loss for learning hierarchical representations,'' in \emph{ICASSP 2022-2022 IEEE International Conference on Acoustics, Speech and Signal Processing (ICASSP)}.\hskip 1em plus 0.5em minus 0.4em\relax IEEE, 2022, pp. 3623--3627.

\bibitem{mcginn2023feature}
K.~McGinn, S.~Kahl, M.~Z. Peery, H.~Klinck, and C.~M. Wood, ``Feature embeddings from the birdnet algorithm provide insights into avian ecology,'' \emph{Ecological Informatics}, vol.~74, p. 101995, 2023.

\bibitem{ghani2023global}
B.~Ghani, T.~Denton, S.~Kahl, and H.~Klinck, ``Global birdsong embeddings enable superior transfer learning for bioacoustic classification,'' \emph{Scientific Reports}, vol.~13, no.~1, p. 22876, 2023.

\bibitem{sethi2020characterizing}
S.~S. Sethi, N.~S. Jones, B.~D. Fulcher, L.~Picinali, D.~J. Clink, H.~Klinck, C.~D.~L. Orme, P.~H. Wrege, and R.~M. Ewers, ``Characterizing soundscapes across diverse ecosystems using a universal acoustic feature set,'' \emph{Proceedings of the National Academy of Sciences}, vol. 117, no.~29, pp. 17\,049--17\,055, 2020.

\bibitem{hershey2017cnn}
S.~Hershey, S.~Chaudhuri, D.~P. Ellis, J.~F. Gemmeke, A.~Jansen, R.~C. Moore, M.~Plakal, D.~Platt, R.~A. Saurous, B.~Seybold \emph{et~al.}, ``Cnn architectures for large-scale audio classification,'' in \emph{2017 ieee international conference on acoustics, speech and signal processing (icassp)}.\hskip 1em plus 0.5em minus 0.4em\relax IEEE, 2017, pp. 131--135.

\bibitem{sarkar23_interspeech}
E.~Sarkar and M.~Magimai.-Doss, ``{Can Self-Supervised Neural Representations Pre-Trained on Human Speech distinguish Animal Callers?}'' in \emph{Proc. INTERSPEECH 2023}, 2023, pp. 1189--1193.

\bibitem{lakdari2024mel}
M.~W. Lakdari, A.~H. Ahmad, S.~Sethi, G.~A. Bohn, and D.~J. Clink, ``Mel-frequency cepstral coefficients outperform embeddings from pre-trained convolutional neural networks under noisy conditions for discrimination tasks of individual gibbons,'' \emph{Ecological Informatics}, vol.~80, p. 102457, 2024.

\bibitem{hagiwara2023beans}
M.~Hagiwara, B.~Hoffman, J.-Y. Liu, M.~Cusimano, F.~Effenberger, and K.~Zacarian, ``Beans: The benchmark of animal sounds,'' in \emph{ICASSP 2023-2023 IEEE International Conference on Acoustics, Speech and Signal Processing (ICASSP)}.\hskip 1em plus 0.5em minus 0.4em\relax IEEE, 2023, pp. 1--5.

\bibitem{hagiwara2023aves}
M.~Hagiwara, ``Aves: Animal vocalization encoder based on self-supervision,'' in \emph{ICASSP 2023-2023 IEEE International Conference on Acoustics, Speech and Signal Processing (ICASSP)}.\hskip 1em plus 0.5em minus 0.4em\relax IEEE, 2023, pp. 1--5.

\bibitem{maaten2008visualizing}
L.~Maaten, ``Visualizing data using t-sne,'' \emph{Journal of machine learning research}, vol.~9, no. Nov, p. 2579, 2008.

\bibitem{mcinnes2018umap}
L.~McInnes, J.~Healy, and J.~Melville, ``Umap: Uniform manifold approximation and projection for dimension reduction,'' \emph{arXiv preprint arXiv:1802.03426}, 2018.

\bibitem{thomas2022practical}
M.~Thomas, F.~H. Jensen, B.~Averly, V.~Demartsev, M.~B. Manser, T.~Sainburg, M.~A. Roch, and A.~Strandburg-Peshkin, ``A practical guide for generating unsupervised, spectrogram-based latent space representations of animal vocalizations,'' \emph{Journal of Animal Ecology}, vol.~91, no.~8, pp. 1567--1581, 2022.

\bibitem{zhang2018automatic}
Y.-J. Zhang, J.-F. Huang, N.~Gong, Z.-H. Ling, and Y.~Hu, ``Automatic detection and classification of marmoset vocalizations using deep and recurrent neural networks,'' \emph{The Journal of the Acoustical Society of America}, vol. 144, no.~1, pp. 478--487, 2018.

\bibitem{padovese2021data}
B.~Padovese, F.~Frazao, O.~S. Kirsebom, and S.~Matwin, ``Data augmentation for the classification of north atlantic right whales upcalls,'' \emph{The Journal of the Acoustical Society of America}, vol. 149, no.~4, pp. 2520--2530, 2021.

\bibitem{nanni2020data}
L.~Nanni, G.~Maguolo, and M.~Paci, ``Data augmentation approaches for improving animal audio classification,'' \emph{Ecological Informatics}, vol.~57, p. 101084, 2020.

\bibitem{yang2021delving}
Y.~Yang, K.~Zha, Y.~Chen, H.~Wang, and D.~Katabi, ``Delving into deep imbalanced regression,'' in \emph{International conference on machine learning}.\hskip 1em plus 0.5em minus 0.4em\relax PMLR, 2021, pp. 11\,842--11\,851.

\bibitem{chen2024neural}
C.~Chen, Z.~Yang, and X.~Wang, ``Neural embeddings rank: Aligning 3d latent dynamics with movements,'' \emph{Advances in Neural Information Processing Systems}, vol.~37, pp. 141\,461--141\,489, 2024.

\end{thebibliography}
\end{document}